\documentclass[a4paper]{jpconf}

\usepackage{graphicx}% Needed for figure files
\usepackage{hyperref}% Gives hyperlinks
\usepackage{amsmath,amsthm,amssymb}% Extra maths stuff

\begin{document}
\title{Lattice-switch Monte Carlo: the fcc--bcc problem}

\author{T L Underwood and G J Ackland}

\address{School of Physics and Astronomy, SUPA, The University of Edinburgh, Edinburgh,
EH9 3JZ, UK }

\begin{abstract}
Lattice-switch Monte Carlo is an efficient method for calculating the free energy difference between two solid phases, or a solid
and a fluid phase.
Here, we provide a brief introduction to the method, and list its applications since its inception. We then describe a lattice 
switch for the fcc and bcc phases based on the Bain orientation relationship. 
Finally, we present preliminary results regarding our application of the method to the fcc and bcc phases in the 
Lennard-Jones system. Our initial calculations reveal that the bcc phase is unstable, quickly degenerating into some
as yet undetermined metastable solid phase. This renders conventional lattice-switch Monte Carlo intractable for this
phase. Possible solutions to this problem are discussed.
\end{abstract}

\section{Introduction}\label{sec:intro}
A common problem which occurs in condensed matter physics is as follows: for a given substance, which of two candidate phases is preferred at a
given, say, temperature $T$ and pressure $P$? This problem amounts to evaluating the free energy difference between the two phases;
the preferred phase has the lower free energy. Unfortunately, calculating the free energy difference between two phases
to a sufficient accuracy to solve this problem can be difficult. This is the case for a plethora of systems of practical 
interest, and is by no means limited to `realistic' models of particle interactions. For instance, while the 
hard-sphere solid is an archetype of a `simple' system, until relatively recently there was contention regarding whether the equilibrium 
phase was fcc or hcp \cite{Bruce_1997}. A similar situation also existed for the Lennard-Jones solid at low temperatures and pressures
\cite{Bruce_2000}.

The method which resolved the aforementioned hard-sphere and Lennard-Jones disputes is \emph{lattice-switch Monte Carlo} (LSMC)
\cite{Bruce_1997,Bruce_2000}. LSMC allows the free energy difference between two phases to be calculated efficiently. 
Furthermore it is `exact' in the sense that it relies upon no approximations other than those present in the model of particle interactions 
it is used in conjunction with. LSMC has been applied to a wide range of systems since its inception -- as summarised in Table \ref{table}%
\footnote{The incarnation of LSMC which can treat solid--\emph{fluid} free energy differences \cite{Wilding_2000} is commonly referred to as 
\emph{phase-switch Monte Carlo}.}.
This reflects the generality of the method; it can in principle be applied to any pair of phases, and any model of particle interactions. 
This feature of LSMC, in combination with its supposedly superior computational efficiency compared to alternative methods, makes it an attractive 
prospect for \emph{ab initio} applications. However, it should be noted that claims of its superiority have proved contentious, and remain 
somewhat of an open question \cite{Pronk_1999,Wilms_2012}.

The layout of this work is as follows. In the following section we provide a brief introduction to LSMC. Detailed accounts of the 
method can be found in the references. 
In Sec. \ref{sec:results} we consider the application of LSMC to the fcc and bcc phases of the Lennard-Jones solid. We first describe a 
lattice switch between these phases, and conclude by presenting the results of our initial investigations.

\begin{table}\label{table}
\caption{Applications of LSMC since its inception. The notation $A \leftrightarrow B$ signifies that LSMC has been used to determine the free
energy difference between phases $A$ and $B$.}
\begin{center}
\begin{tabular}{lll}
\br
Phases considered & Interparticle potential & References \\
\mr
fcc $\leftrightarrow$ hcp & Hard sphere & \cite{Bruce_1997,Bruce_2000,Pronk_1999} \\
Various fcc and hcp phases with stacking faults & Hard sphere & \cite{Pronk_1999,Mau_1999} \\
fcc $\leftrightarrow$ fluid & Hard sphere & \cite{Wilding_2000} \\
fcc $\leftrightarrow$ hcp & Lennard-Jones & \cite{Jackson_2002} \\
square $\leftrightarrow$ triangular & Core-softened & \cite{Wilding_2002} \\
fcc $\leftrightarrow$ fluid & Lennard-Jones & \cite{Errington_2004,McNeil-Watson_2006} \\
Various close-packed polydisperse binary phases & Hard sphere & \cite{Jackson_2007} \\
fcc $\leftrightarrow$ hcp (both polydisperse) & Hard sphere & \cite{Yang_2008} \\
fcc $\leftrightarrow$ fluid, bcc$\leftrightarrow$fluid & r-12 & \cite{Wilding_2009_MP} \\
fcc $\leftrightarrow$ fluid (both polydisperse)& r-12 & \cite{Wilding_2009_JCP,Sollich_2010,Wilding_2010} \\
triangular $\leftrightarrow$ triangular `soliton staircase' & r-12 & \cite{Wilms_2012} \\
\br
\end{tabular}
\end{center}
\end{table}

\section{Lattice-switch Monte Carlo: an introduction}
Consider the Gibbs free energy difference $\Delta F\equiv F_1-F_2$ between two crystalline phases 1 and 2, where $F_{\alpha}$ denotes the
free energy of phase $\alpha$. It can be shown that
\begin{equation}\label{DeltaF_useful}
\Delta F=\beta^{-1}\ln\big(\,p_2/p_1\bigr),
\end{equation}
where $\beta$ denotes the thermodynamic beta, and $p_{\alpha}$ denotes the probability of the system being in phase $\alpha$ at thermodynamic equilibrium
-- assuming that the system is constrained to be in either phase 1 or phase 2.
The above equation can be exploited to calculate $\Delta F$ theoretically: extract $p_2/p_1$, the time the system spends in phase 2 relative
to phase 1, from an $NPT$ simulation of the system, and substitute this quantity into the above equation. 
However, this approach is intractable for simulations utilising `realistic' particle dynamics if the typical time taken for the system to 
transition between the two phases is very long, in which case a reasonable estimate of $p_2/p_1$ cannot be deduced in a reasonable simulation time. It may 
even be the case that, regardless of the phase in which the simulation is initialised, the system \emph{never} transitions to the `other' phase 
during the course of the simulation.
This stems from the fact that, at equilibrium, the most probable states comprise two `islands of stability' in phase space: 
one within phase 1 and the other within phase 2. However, these two islands are separated by an \emph{entropic barrier}: a region of phase space 
comprised of states which are very improbable at equilibrium. Hence to transition between the phases the system must traverse the
entropic barrier, the success of which is very unlikely, and hence a rare occurrence.

The traditional implementation of Metropolis Monte Carlo \cite{Metropolis_1953} belongs to the aforementioned class of simulations which utilise
`realistic' particle dynamics. In this approach (for an $NPT$ simulation) at each time-step a trial state of the system is generated 
from the current state
by either altering the position of one of the particles, i.e., a `particle move' is attempted, or altering the dimensions of the simulation box, i.e.,
a `volume move' is attempted. The trial state is then accepted or rejected as the new state of the system for the next time-step according to the Metropolis 
algorithm \cite{Metropolis_1953}. The end result is that for a long simulation the states are sampled from the probability distribution corresponding to 
thermodynamic equilibrium.
However, the important properties of the Metropolis algorithm do not rely upon the mechanism to generate trial states just described;
one has considerable freedom with regards to how trial states are generated. 
The prospect therefore exists of generating states in a manner such that the system 
traverses a path in phase space which allows $\Delta F$ to be accurately calculated in relatively few time-steps. Such a path would involve frequent 
transitions between both phases by `jumping over' the entropic barrier. 

This is what is done in LSMC; a new type of move, a \emph{lattice switch}, is introduced in order to supplement the aforementioned particle and
volume moves.
Intuitively, the system is in phase $\alpha$ if the positions of the particles approximately form the crystal lattice characteristic of $\alpha$ at the
current volume. With this in mind, if the system is in phase $\alpha$, one can express the position $\mathbf{r}_i$ of particle $i$ as
\begin{equation}
\mathbf{r}_i=\mathbf{R}_{\alpha,i}+\mathbf{u}_i,
\end{equation}
where $\mathbf{R}_{\alpha,i}$ denotes the position of the $\alpha$ crystal lattice site which is closest to $i$, and $\mathbf{u}_i$ denotes 
the displacement of $i$ from that lattice site.
The trial state $\sigma'$ generated by a lattice switch from a state $\sigma$ in phase 1 shares the same set of particle displacements 
$\lbrace\mathbf{u}_i\rbrace$ as $\sigma$, but the underlying set of lattice vectors for $\sigma'$ is $\lbrace\mathbf{R}_{2,i}\rbrace$ instead of 
$\lbrace\mathbf{R}_{1,i}\rbrace$; the underlying lattice is `switched' from $\sigma$ to $\sigma'$. In other words, the lattice switch amounts to the 
following transformation for all $i$:
\begin{equation}
\mathbf{r}_i=\mathbf{R}_{1,i}+\mathbf{u}_i\to\mathbf{R}_{2,i}+\mathbf{u}_i.
\end{equation}
Hence every time a lattice switch is accepted the system transitions directly to the `other' phase, bypassing the entropic barrier.

This is, however, only half of the story. One might expect that by regularly attempting lattice switches the system will frequently transition between 
phases, allowing $\Delta F$ to be efficiently evaluated as described above. However, if one does this with Metropolis Monte Carlo, one 
finds that lattice switches are too rarely accepted for this approach to be useful. 
The reason for this is that, for the states visited during a typical simulation, $\sigma'$ is almost always of a much higher energy than $\sigma$,
and hence lattice switches will almost always be rejected by the Metropolis algorithm.
Crucially, there exist states for which $\sigma$ and $\sigma'$ are of comparable energy; from such states switches have a reasonable chance of success. 
We refer to these states as \emph{gateway states}, since they provide the key to jumping between both phases. 
Unfortunately, these states are almost never visited
dduring a Metropolis Monte Carlo simulation. To encourage successful lattice switches, we therefore use \emph{multicanonical Monte Carlo}
\cite{Berg_1991,Berg_1992,Smith_1995} -- which provides a means of sampling selected states more (or less) frequently than is the case at 
equilibrium, while still allowing equilibrium properties of the system to be evaluated -- to more frequently visit gateway states.
The result is that lattice switches are accepted reasonably often, and both phases are explored in a reasonable simulation time.
To elaborate, we introduce a quantity $\mathcal{M}$ which characterises how `gateway-like' a state is, with 
$\mathcal{M}=0$ corresponding to `perfectly gateway-like', and $|\mathcal{M}|\gg 0$ corresponding to `very un-gateway-like'.
The specific definition of the quantity $\mathcal{M}$ depends on the system under consideration. For solid--solid free energy
differences in soft-potential systems, the following definition of $\mathcal{M}$ has been used:
\begin{equation}
\mathcal{M}(\lbrace\mathbf{u}_i\rbrace)=E(\lbrace\mathbf{R}_{1,i}+\mathbf{u}_i\rbrace)
-E(\lbrace\mathbf{R}_{2,i}+\mathbf{u}_i\rbrace),
\end{equation}
where $E(\lbrace\mathbf{r}_i\rbrace)$ denotes the energy of the state with particle positions $\lbrace\mathbf{r}_i\rbrace$. The first term on the 
right-hand side is the energy associated with the displacements $\lbrace\mathbf{u}_i\rbrace$ for phase 1, and the second term is the analogous 
quantity for phase 2.
Note that $\mathcal{M}(\lbrace\mathbf{u}_i\rbrace)=0$ if the energies associated with $\lbrace\mathbf{u}_i\rbrace$ in both phases are identical. In this
case the energy cost of a lattice switch from either phase is zero, and hence the states associated with $\lbrace\mathbf{u}_i\rbrace$ in both phases 
are gateway states.
Defining a macrostate as a collection of all states with the same $\mathcal{M}$, we then sample all macrostates in our multicanonical
simulation with equal probability. The result is that the gateway-like macrostates are frequently visited.

\section{The fcc--bcc transition in the Lennard-Jones solid}\label{sec:results}
We now turn to the problem of using LSMC to evaluate the free energy difference between the fcc and bcc phases in the Lennard-Jones system. Our motivation
behind this is twofold. Firstly, the fcc--bcc transition is of profound importance to metallurgy. Our ultimate aim is to apply LSMC to
this transition using more realistic models of metals than the Lennard-Jones model, such as the embedded atom model \cite{Daw_1984}, 
or even \emph{ab initio} models. 
However, given that there has yet to be a LSMC study of the fcc--bcc transition, it is sensible to `tread carefully' and first study the fcc--bcc
transition using the simpler, and better understood, Lennard-Jones model before proceeding to more uncharted waters. 
Secondly, there has been speculation that there is a region in the phase diagram of the Lennard-Jones system at high temperatures and pressures, 
below the melting curve, where the bcc phase will be preferred over the fcc \cite{Rahman_1984}. It would be interesting to test this hypothesis, the 
confirmation of which would have far-reaching consequences given the widespread use of the Lennard-Jones model to describe real systems.

\subsection{The fcc--bcc lattice switch}
\begin{figure}
\centering
\includegraphics[width=0.8\textwidth]{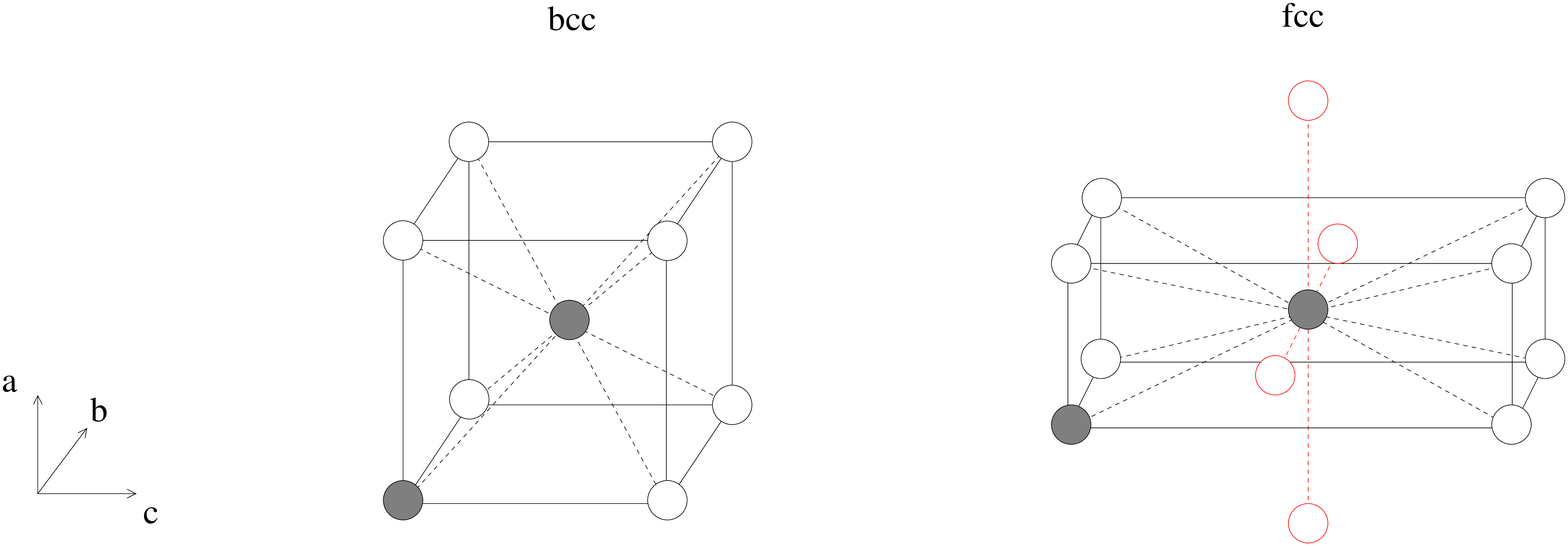}
\caption
{(Colour online) Schematic diagram illustrating the bct representations of the bcc and fcc crystal structures. Each unit cell contains two particles,
which are represented by grey circles. Image particles are represented by white circles. Dashed lines connect the particle at the centre of each unit
cell to its nearest neighbours. In the fcc case, four of the twelve nearest neighbours lie beyond the edges of the unit cell. These neighbours and the
corresponding dashed lines are coloured red.}
\label{fig:fcc_bcc_switch}
\end{figure}

After deciding to apply LSMC to a certain system, the first problem one encounters is the choice of lattice switch.
A lattice switch is a one-to-one mapping of particle positions in one phase to another phase. 
Hence it is necessary that the supercells used to represent both phases
have the same number of particles. The Bain orientation relationship (see, e.g., Ref. \cite{book:Mittemeijer}) provides a means for achieving
this for the case of the fcc and bcc phases. To elaborate, both fcc and bcc crystals can be recast as bct crystals: the bcc crystal is equivalent to
a bct crystal in which the bct unit cell has equal dimensions, i.e. $a=b=c$; the fcc crystal is equivalent to a bct crystal in which
$a=b=c/\sqrt 2$. This is illustrated in Fig. \ref{fig:fcc_bcc_switch}.
Thus by tiling $N_a$, $N_b$ and $N_c$ bct unit cells corresponding to fcc or bcc along the $a$-, $b$- and $c$-directions, one can 
construct a bcc and a fcc supercell which both contain the same number of lattice sites $2N_aN_bN_c$. Specifically, the positions of the lattice sites 
in one of the supercells are given by $(n_aa,n_bb,n_cc)$ and $(n_aa+a/2,n_bb+b/2,n_cc+c/2)$ where $n_a=0,1,\dotsc,(N_a-1)$ and 
similarly for $n_b$ and $n_c$.

We have just described how to construct an fcc and a bcc supercell with the same number of lattice sites. For the $NVT$ ensemble we are interested in 
fcc and bcc phases with the same density $\rho$, and the supercells should reflect this. It can be shown that $a_{\text{bcc}}=2^{1/3}\rho^{-1/3}$, 
$a_{\,\text{fcc}}=2^{1/6}\rho^{-1/3}$ and $c_{\,\text{fcc}}=2^{2/3}\rho^{-1/3}$. Hence for fcc and bcc crystals of equal density it is necessary that 
$a_{\,\text{fcc}}=2^{-1/6}a_{\text{bcc}}$ and $c_{\,\text{fcc}}=2^{1/3}a_{\text{bcc}}=2^{1/3}c_{\text{bcc}}$. Therefore, starting from the bcc supercell, if one applies the 
transformation
\begin{equation}
a\to 2^{-1/6}a, \quad b\to 2^{-1/6}b, \quad c\to 2^{1/3}c
\end{equation}
to the lattice site positions and the dimensions of the supercell, then the end result is an fcc supercell with the same density. This corresponds
to stretching the bcc supercell in the $c$-direction, while simultaneously compressing the supercell in the $a$- and $b$-directions to preserve its
volume. Taking this idea further, if one makes the aforementioned transformation, but keeps the displacements $\lbrace \mathbf{u}_i\rbrace$ of the particles 
in the supercell unchanged, then one has performed a density-preserving lattice switch from bcc to fcc. Obviously the converse operation is a 
density-preserving lattice switch from fcc to bcc.
In an $NPT$ ensemble it may be the case that, say, the bcc phase is of a lower density than the fcc phase, in which case a switch which 
increases the density upon transforming from bcc to fcc, and correspondingly lowers the density upon transforming from fcc to bcc, may yield a more 
efficient simulation. The above discussion can be easily adapted to treat such non-density-preserving lattice switches.

\subsection{Results of initial investigations: instability of the bcc phase}
\begin{figure}
\centering
\includegraphics[height=0.8\textwidth,angle=270]{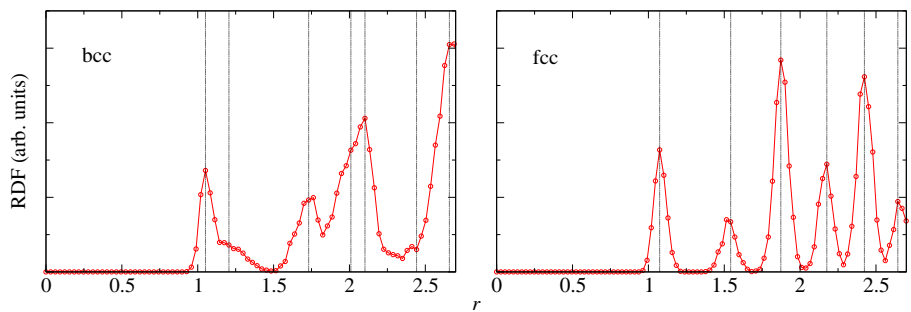}
\caption
{(Colour online) The radial distribution functions obtained at the completion of simulations of the Lennard-Jones system initialised in the bcc and 
fcc phases -- as described in the main text. The red curve and circles in the left(right) panel is the RDF corresponding to the simulation initialised in 
the bcc(fcc) phase. $r$ denotes the distance from a particle, in reduced units. The dashed, black vertical lines indicate the locations of the `bins' of 
the RDF histogram which have non-zero ordinates for a perfect bcc or fcc crystal. In other words, in each panel the $n$th line from the left corresponds
to the $n$th nearest neighbour shell in the analogous perfect crystal.}
\label{fig:rdf}
\end{figure}

Finally we turn to our simulations, the methodology of which closely resembles that of Ref. \cite{Jackson_2002}. Before performing a
LSMC simulation to calculate $\Delta F$ to a high degree of accuracy,
one must optimise the step size used in the particle and volume moves used to generate trial states each time step. To do this it is sufficient to
use Metropolis Monte Carlo simulations locked into one of the phases, with a small system size. Furthermore, such simulations act as a `sanity 
check' before more accurate simulations are undertaken. It was during such simulations that we noticed that the bcc phase would quickly degenerate
into some other -- as yet unidentified -- metastable phase. The same was not observed to occur for the fcc phase. This is illustrated in Fig.
\ref{fig:rdf}, which shows the radial distribution functions (RDFs) at the end of $NVT$ Metropolis Monte Carlo simulations 
of systems of 250 particles at $\rho\approx 1.1$ and $\beta=3.333$ initialised in the bcc and fcc phases, where we are using reduced units as 
described in Ref. \cite{Jackson_2002}. Each simulation comprised $1\times 10^7$ particle moves, and repeats of the simulations yielded very 
similar RDFs to those shown in the figure. Note that the peaks in 
the fcc RDF are in excellent agreement with those of the perfect fcc crystal; however, the same cannot be said for the bcc RDF.

The fact that the bcc phase is so short-lived makes it impossible to apply LSMC as it stands to determine the free energy between fcc and
bcc in the Lennard-Jones system. A similar problem was described in Ref. \cite{Wilding_2002} for two-dimensional core-softened systems.
It should be borne in mind that if one considers \emph{any} two phases, at least one of them will be metastable, and hence will destabilise
given a long enough simulation time; we require that the two phases under consideration do not destabilise before
the simulation time required to determine $\Delta F$ to the desired accuracy is reached. With regards to the fcc--bcc problem, the following
question comes to mind: is there a way in which the system can be kept in the bcc phase for long enough to gather decent statistics relevant to 
calculating $\Delta F$, or indeed any other property? One might think that a particle move mechanism which constrains the system
to remain within the phase under consideration is a valid means of preventing the bcc phase from destabilising. 
However, `hard wall' constraints on the particle positions can lead to `drift' in the 
centre-of-mass of the simulation, which may invalidate the final results \cite{Bruce_2000}. 
An alternative approach is to softly `tether' particles to their lattice sites in the multicanonical simulation through judicious choice
of the weight function. A similar idea is used in applying LSMC to fluids \cite{Wilding_2000}, 
and may be worth exploring as a means of addressing instability in solids.

\ack
This work was supported by the Engineering and Physical Sciences Research Council.

\section*{References}

%% The bibliography. Comment this out and insert the .bbl file after it for the final copy - so that only one .tex file is needed.
\bibliographystyle{iopart-num}
\bibliography{bibliography}

\end{document}